\documentclass[twocolumn,prb,showpacs,floatfix]{revtex4}
\usepackage{latexsym}
\usepackage{graphicx}
\usepackage{amssymb}
\usepackage{amsmath}

\begin{document}

\title{Admixtures to $d$-wave gap symmetry in untwinned YBa$_{2}$Cu$_{3}$O$_{7}$
superconducting films measured by angle-resolved electron tunneling}
\author{H.J.H. Smilde, A.A. Golubov, Ariando, G. Rijnders, J.M. Dekkers, S.
Harkema, D.H.A. Blank, H. Rogalla, and H. Hilgenkamp}
\affiliation{Faculty of Science and Technology and MESA$^{+}$
Research Institute, University of Twente, P.O. Box 217, 7500 AE Enschede,
The Netherlands}

\date{\today}

\begin{abstract}
We report on an \textit{ab}-anisotropy of $J_{c \parallel b}/J_{c \parallel a}%
\cong 1.8$ and $I_{c}R_{n \parallel b}/I_{c}R_{n \parallel a}\cong 1.2$ in
ramp-edge junctions between untwinned YBa$_{2}$Cu$_{3}$O$_{7}$ and $s$%
-wave Nb. For these junctions, the angle $\theta $ with the
YBa$_{2}$Cu$_{3}$O$_{7}$ crystal $b$-axis is varied as a single
parameter. The $R_{n}$A($\theta$)-dependence presents 2-fold symmetry. The
minima in $I_{c}R_{n}$ at $\theta $~$\cong $~$50^{\circ}$ suggest a real $s$-wave
subdominant component and negligible $d_{xy}$-wave or imaginary $s$-wave
admixtures. The $I_{c}R_{n}$($\theta$)-dependence is well-fitted by $83\%$
$d_{x^{2}-y^{2}}$-, $15\%$ isotropic $s$- and $2\%$ anisotropic $s$-wave
order parameter symmetry, consistent with $\Delta _{b}/\Delta _{a} \cong 1.5$.
\end{abstract}

\maketitle

Phase-sensitive experiments\cite{Wollmann:1993,Tsuei:1994} and
tunnel spectroscopy\cite{Wei:1998} have provided rich evidence for the sign
change of the pair wave function in the crystal $ab$-plane of
high-$T_{c}$ superconductors.
Insight in the extent of subdominant admixtures to the $d_{x^{2}-y^{2}}$%
-wave symmetry is less well established. They are of high importance for the
basic understanding of high-$T_{c}$ superconductivity and the design of
novel $d$-wave based Josephson devices,
but also for standard high-$T_{c}$ junctions. They determine for
instance the exact position of the nodes, and the amount of $ab$-anisotropy.

In YBa$_{2}$Cu$_{3}$O$_{7}$ strong anisotropy in the electronic
structure has been reported, which can be interpreted as an
effective mass anisotropy along the $a$- and $b$-axes: An elongated
vortex shape by scanning tunneling spectroscopy \cite{Maggio:1995}
suggests 50$\%$ anisotropy. Sixty percent anisotropy is found in the
London penetration depth by far-infrared
spectroscopy,\cite{Basov:1995} as well as using $c$-axis
YBa$_{2}$Cu$_{3}$O$_{7}$/Pb Josephson junctions with a magnetic
field oriented parallel to the $a$- or $b$-axis.\cite{Sun:1995}
Other studies, neutron scattering on flux-line lattices
\cite{Johnson:1999} and single crystal
torque-measurements,\cite{Ishida:1996} indicate a smaller anisotropy
of $1.2$. Related, surface impedance \cite{Zhang:1994} and
resistivity measurements \cite{Friedmann:1990} demonstrate an
anisotropy of $R_{s\parallel a}/R_{s\parallel b}\approx 1.5$ to
$1.6$ and $\sqrt{\rho _{a}/\rho _{b}}\approx 1.5$ respectively.

Also, implications for the anisotropy of the superconducting gap
have been discussed. Raman scattering \cite{Limonov:1998} evidences
a real isotropic $s$-wave admixture of 5$\%$; thermal conductivity
measurements in a rotating magnetic field \cite{Aubin:1997} place a
maximum of 10{\%} based on the node positions. Angle-resolved
photoemission spectroscopy (ARPES)\cite{Aet:2001} indicates larger
$ab$-anisotropy of $\Delta _{b}/\Delta _{a}=1.5$. The use of
untwinned single-crystals is considered crucial in all these
studies. However, clear consensus on subdominant order parameter
symmetries is not reached, nor is detailed angle-resolved data in
the $ab$-plane of thin films available, although first attempts on
twinned films have been performed.\cite{van:1999} In view of this,
we present here new results on the anisotropy, comparing untwinned
and twinned YBa$_{2}$Cu$_{3}$O$_{7}$.

\begin{figure}[bt]
\centerline{\includegraphics[width=2.5in]{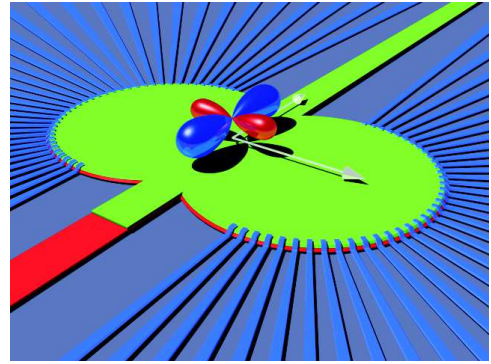}}
\caption{(Color online) Angle-resolved electron tunneling with YBa$_{2}$Cu$_{3}$O$_{7}$%
/Au/Nb ramp-type junctions oriented every $5^{\circ}$ over $360^{\circ}$. The YBa$_{2}$%
Cu$_{3}$O$_{7}$ base-electrode (red) is covered by SrTiO$_{3}$ insulator
(green), and contacted by a Au barrier (not visible) and a Nb
counter-electrode (blue). The arrows (white) indicate the main crystal
orientations in the $ab$-plane of the high-$T_{c}$
superconducting material.}
\label{fig1}
\end{figure}

In untwinned YBa$_{2}$Cu$_{3}$O$_{7}$ thin films, the usual `random'
exchange of the $a$- and $b$-axis is eliminated. This enables to
study the electronic properties angle-resolved in the $ab$-plane. The
experimental layout is summarized in Fig.~\ref{fig1}. Basically, the YBa$_{2}$Cu$_{3}$%
O$_{7}$ base-electrode is patterned into a nearly circular polygon, changing
the orientation from side to side by 5 degrees. A Au barrier and Nb
counter-electrode contact each side. In this way, the angle with respect to
the $(0\!\; 1\!\; 0)$-orientation is varied as a single parameter.

First, bilayers of 170~nm YBa$_{2}$Cu$_{3}$O$_{7}$ and 100~nm SrTiO$_{3}$ are grown by
pulsed-laser deposition (PLD) on single-crystal SrTiO$_{3}$ substrates. The
YBa$_{2}$Cu$_{3}$O$_{7}$ films are optimally doped, with $%
T_{c,0}$~$\geq $~89~K. Ramps are ion-milled in the bilayers using a photo
resist stencil. To assure equivalent ramp quality over $360^{\circ}$, the sample
stage is rotated around the substrate normal, while maintaining the angle of
incidence of the Ar-ion beam constant at $40^{\circ}$ with the substrate plane.
The resulting ramp-angle with the substrate plane is $\alpha _{R}$~$\cong $~$%
30^{\circ}$. On a microscopic scale the interfaces may present some faceting,
albeit less than in e.g. grain boundaries. This faceting is not expected to
affect the main conclusions of the presented studies. After removal of the
photo-resist stencil and a short $90^{\circ}$%
-incidence ion-mill for cleaning purposes, a 5~nm YBa$_{2}$Cu$_{3}$O$_{7}$
interlayer \cite{Smilde:2002} is deposited to prepare an in-situ interface
to a 30~nm Au barrier formed also by PLD. Then, a 160~nm thick Nb
counter-electrode is dc-sputter deposited through a lift-off stencil. Special care
is taken to obtain a clean Au/Nb interface by a 50~s rf-plasma etch just before Nb
deposition. After lift-off, redundant Au and YBa$_{2}$Cu$_{3}$O$_{7}$
interlayer material is removed by Ar-ion milling. The junctions are 4~$\mu $%
m wide.

\begin{figure}[tb]
\centerline{\includegraphics[width=3in]{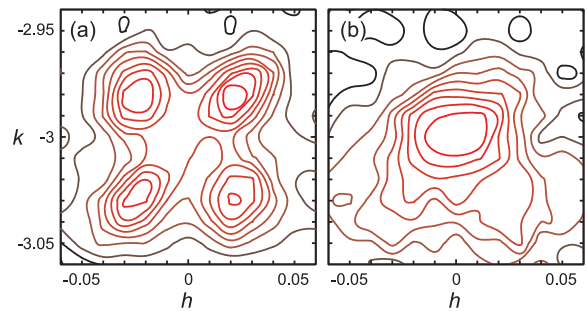}}
\caption{(Color online) Logarithmic contour plots of $hk$-scans near the $(0\!\; %
\overline 3\!\; 4)$ reflection of YBa$_{2}$Cu$_{3}$O$_{7}$: (a) grown on a
small vicinal angle SrTiO$_{3} $ substrate (\textit{$\protect\alpha $}~$\approx
$~$0.12^{\circ}$), and (b) grown on an \textit{$\protect\alpha $}~$\protect\cong %
$~$1.07^{\circ}$ and \textit{$\protect\beta $}~$\protect\cong$~$-2.1^{\circ}$ vicinal SrTiO%
$_{3}$ substrate. Both scans are measured after device completion.}
\label{fig2}
\end{figure}

The twin behavior of $(0\!\; 0\!\; 1)$-YBa$_{2}$Cu$_{3}$O$_{7}$ films is influenced
by the substrate vicinal angle $\alpha $ and its in-plane orientation $\beta $%
.\cite{Dekkers:2003} Here, $\alpha $ is defined between the crystallographic
and optical substrate-normal, and $\beta $ describes the in-plane
orientation with respect to the SrTiO$_{3}$ $\langle 1\!\; 0\!\; 0\rangle $ crystal
axis. The degree of twinning can be controlled from completely untwinned to the
presence of four $ab$-orientations, varying $\alpha $ from $\sim $~$%
1.1^{\circ}$ to a small vicinal angle ($\sim $~$0.1^{\circ}$), where $\beta $~$\cong $~%
$0^{\circ}$. For $\alpha $~$\cong $~$1.1^{\circ}$, growth with the $b$-axis
along the step-ledges is induced, and only one crystal orientation is
present. On the contrary, 4 twin orientations are present for small vicinal angle
substrates. The twin orientations have pair-wise their in-plane diagonal of
the YBa$_{2}$Cu$_{3}$O$_{7}$ crystal aligned with each substrate diagonal,
so that $a$- and $b$-axes and vice versa are arranged nearly in
parallel.\cite{Dekkers:2003}  After completion of the device, the YBa$_{2}$Cu$_{3}$O$_{7}$
base-electrode is examined with X-ray diffraction (XRD). An average of
the $a$ and $b$ unit cell dimensions is found for twinned films (see
Table~\ref{table1}). For untwinned films, the individual $a$ and $b$ unit cell
parameters can be distinguished and are close to single-crystal values.
Detailed $hk$-scans of the $(0\!\; \overline{3}\!\; 4)$ reflections show
4 different orientations for YBa$_{2}$Cu$_{3}$O$_{7}$ films grown on
small vicinal angle substrates (Fig.~\ref{fig2}a), associated with the above-mentioned
4 twin orientations. For films grown on substrates with $\alpha $%
~$\cong $~$1.1^{\circ}$ however, only one orientation is present (Fig.~\ref{fig2}b).

\begin{table}[tb]
\begin{tabular}{r c r @{.} l @{/} c @{} r @{.} l r @{.} l @{} c @{/} r @{.} l}
\hline
& & \multicolumn{5}{c}{twinned} & \multicolumn{5}{c}{untwinned} \\ \hline
SrTiO$_{3}$ : $\alpha$/$\beta$ & \ \ & $0$&$12^{\circ}$& &$119$&$0^{\circ}$ & $1$&$07^{\circ}$& &$357$&$9^{\circ}$
\\
YBa$_{2}$Cu$_{3}$O$_{7}$ : $\alpha $/$\beta $ & \ \ & $0$&$20^{\circ}$& &$97$&$3^{{\circ}}$ & $%
0$&$75^{\circ}$& &$~346$&$3^{\circ}$ \\
$a$& \ \ & $3$&\multicolumn{4}{l}{\!\!\!\!\!\; $866(3)$} & $3$&\multicolumn{4}{l}{\!\!\!\!\!\; $849(6)$}
\\
Cell Par. (\AA ) \ \ $b$& \ \ & $3$&\multicolumn{4}{l}{\!\!\!\!\!\; $867(4)$} & $3$&\multicolumn{4}{l}{\!\!\!\!\!\; $892(7)$} \\
$c$& \ \ & $11$&\multicolumn{4}{l}{\!\!\!\!\!\; $678(3)$} & $11$&\multicolumn{4}{l}{\!\!\!\!\!\; $703(7)$}
\\ \hline
\end{tabular}%
\caption{ Vicinal angle $\protect\alpha $ and its orientation $\protect%
\beta $ of the SrTiO$_{3}$ substrates and YBa$_{2}$Cu$_{3}$O$_{7}$ films.
The YBa$_{2}$Cu$_{3}$O$_{7}$ data is obtained after device completion.}
\label{table1}
\end{table}

The {XRD} and electrical data presented in this article correspond to the
same samples. Figure~3 presents the electrical characterization of the
twinned base-electrode sample (a-c), and the untwinned one
(d-f,h). During the measurement, the magnetically shielded sample space
reduces background fields below 0.1~$\mu T$. Trapped flux in or near the
junctions is excluded by systematic $I_{c}(B)$ measurements, assuring a
correctly determined critical current density ($J_{c}$). The superconducting
properties of the Au/Nb bilayer are independent of the orientation. Therefore,
$J_{c}$ depends on the in-plane orientation $\theta $ with respect to the
$b$-axis of the YBa$_{2}$Cu$_{3}$O$_{7}$ crystal only, and presents four
maxima for both samples, approaching zero in between. This is in
agreement with predominant $d_{x^{2}-y^{2}}$-wave symmetry of the
superconducting wave function in one electrode only, and a $\cos(2\theta )$%
-dependence is expected.\cite{Sigrist:1992} In closer detail, the nodes of
the untwinned YBa$_{2}$Cu$_{3}$O$_{7}$ sample are found at $5^{\circ}$
from the diagonal between the $a$- and $b$-axis. This presents direct
evidence for a significant real isotropic $s$-wave admixture. A
first estimate for the $s$- over $d_{x^{2}-y^{2}}$-wave
gap-ratio is calculated as $\vert \cos(2\theta _{0}) \vert \cong 17\%$
for a node angle $\theta _{0}=50^{\circ}$. For the twinned base-electrode,
the nodes are found at the diagonal, which is expected if all twin
orientations are equally present, and contributions of subdominant
components average out to zero.

\begin{figure*}[tb]
\centerline{\includegraphics[width=6.0in]{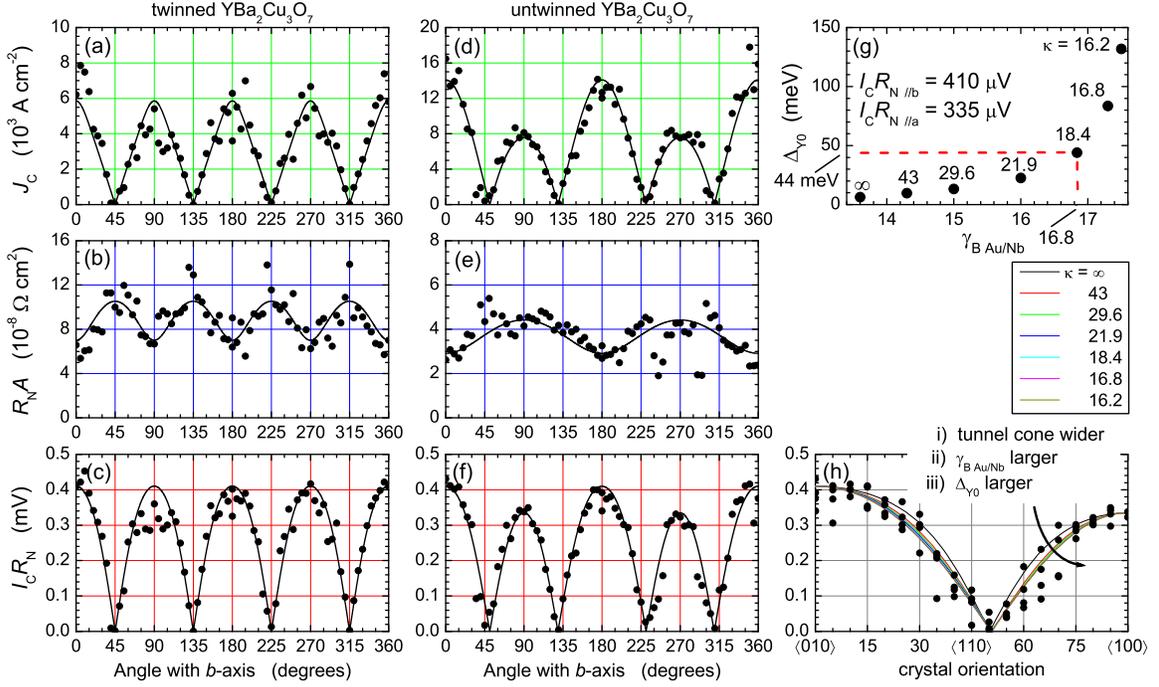}} \caption{(Color
online) $J_{c}$, $R_{n}A$, and $I_{c}R_{n}$-product \textit{vs} the
junction orientation with respect to the YBa$_{2}$Cu$_{3}$O$_{7}$
crystal for twinned (a-c), and untwinned (d-f,h)
YBa$_{2}$Cu$_{3}$O$_{7}$ at $T$~=~4.2~K and in zero magnetic field.
(see text for description fits) (g) Fit parameter evolution for
untwinned case, and (h) corresponding fits.} \label{fig3}
\end{figure*}

The suppressed $J_{c}$ in the nodal direction ($\leq
$~0.01~$J_{c\parallel \langle 0\!\; 1\!\; 0\rangle })$ suggests
small, if not absent, imaginary admixtures, \cite{van:1995} for
instance of isotropic $is$-wave or $id_{xy}$-wave type which in
contrast would lift the nodes. A significant real $d_{xy}$-wave
admixture is excluded, because this would induce a rotation in the
same direction with respect to the crystal of all nodes.

In the untwinned case, the $J_{c}$-value is 1.8 times larger in the
$b$- than in the $a$-direction. Preparation effects can be
eliminated, since circular symmetry with respect to the substrate normal has
been conserved at all phases of the fabrication. The normal-state resistance
($R_{n}A$) is lower along
the $b$- than along the $a$-axis, and presents a two-fold symmetry
axis for the untwinned case. Using the angle-resolved values, the
anisotropy in the $I_{c}R_{n}$-product amounts to $I_{c}R_{n\parallel
b}/I_{c}R_{n\parallel a}\cong 1.22$.

To estimate the $I_{c}R_{n}$-products in our junctions, we
model them as SINS' structures, where S is YBa$_{2}$Cu$_{3}$O$_{7}$,
I is the YBa$_{2}$Cu$_{3}$O$_{7}$/Au interface barrier, which has
much higher resistance than the Au (N) and the Au/Nb (N/S')
interface. From independent resistance measurements on our PLD Au ($\rho
_{Au}$~$\cong $~4.6~$\mu \Omega $~cm at 4.2~K), the
mean-free-path is $l_{Au}$~$\cong $~18~nm, and the dirty-limit
coherence length is $\xi _{Au}$~$\cong $~49~nm. Using these
values and the YBa$_{2}$Cu$_{3}$O$_{7}$/Au interface resistance
$R_{B}A_{Y/Au}$~$\geq $~10$^{-8}$~$\Omega $~cm$^{2}$, the transparency at this
interface is low: $\gamma _{B_{Y/Au}}\geq 440$,
where $\gamma _{B}=R_{B}A/\rho _{Au}\xi _{Au}$.\cite{Golubov:1995}
From a small Fermi-velocity mismatch, we estimate the Au/Nb interface
transparency much larger, $\gamma _{B_{Au/Nb}}$~$<$~20. The
electrode-separation is $d_{Au}=26$~nm for $30$~nm thick Au
and ramp angle $\alpha _{R}~=~30^{\circ}$. Since $l_{Au}$~$<$%
~$d_{Au}$ and $l_{Au}<\xi _{Au}$, Au is in the
diffusive regime, while YBa$_{2}$Cu$_{3}$O$_{7}$ is in
the clean limit with the anisotropic gap function $\Delta _{Y}$.

We extend the expression for the supercurrent in diffusive SINS' structures
\cite{Golubov:1995} to our case of a low-transparent junction
between a clean $d$-wave superconductor and a diffusive NS' bilayer.
The contribution of midgap Andreev bound states is small in such
a junction \cite{Riedel:1998} and can be neglected.
\begin{equation}
I_{s}R_{n}=\frac{1}{N}\iint d\phi d\chi ~\sin(\chi ) ~D ~\gamma ~\sin(\Delta \varphi )
\end{equation}%
\begin{equation}
\gamma =~\frac{2\pi k_{B}T}{e}\sum_{n=0}^{\infty }~\frac{\Phi \Delta _{Y}~}{%
\sqrt{\Phi ^{2}+\omega _{n}^{2}}\sqrt{\Delta _{Y}^{2}+\omega _{n}^{2}}}
\end{equation}%
\begin{equation}
\Phi =\frac{\pi k_{B}T_{cNb} \Delta _{Nb}}{\pi k_{B}T_{cNb}~+~\gamma
_{B_{Au/Nb}}~(d_{Au}/\xi _{Au})~\sqrt{ \Delta _{Nb}^{2}+\omega
_{n}^{2}}}
\end{equation}%
Here $\chi $ is the angle with the interface normal, and $ N=\iint
d\phi d\chi \sin(\chi )D$ is the normalization constant. The
integration is performed over angles $\phi = 0$ to $2\pi $, and
$\chi = 0$ to $\frac{\pi}{2}$ of a half-sphere of all trajectories:
for each junction orientation, and taking the crystal orientation
and ramp-angle into account. The barrier transmission coefficient
$D=\cos(\chi)~\exp [\kappa \{1-\cos^{-1}(\chi )\}]$ is in the limit
of a small YBa$_{2}$Cu$_{3}$O$_{7}$ Fermi velocity, where $\kappa $
describes the tunnel-cone size. The sum in Eq.(2,3) is taken over
the Matsubara frequencies $\omega _{n} = \pi T(2n+1)$. $\Delta _{Y}$
is the anisotropic gap function in YBa$_{2}$Cu$_{3}$O$_{7}$, and
$\Phi $ is the isotropic proximity-induced gap function in Au.$\
\Delta _{Nb}$ and $T_{cNb}$ are the bulk pair potential and the
critical temperature of Nb, respectively. The critical current
$I_{c}$ and $I_{c}R_{n}$-product should be found by calculating a
maximum of $I_{s}R_{n}$ over the phase difference $\Delta \varphi $
across the junction.

Tunneling along the $a$- and $b$-axis may then be compared
theoretically in terms of the ratio $\Gamma =I_{c}R_{n\parallel
b}/I_{c}R_{n\parallel a}$. Using Eqs.(1-3), it can be shown that for
constant properties of the Nb/Au bilayer, the ratio of the
YBa$_{2}$Cu$_{3}$O$_{7}$ gap for these directions is $\Delta
_{Y\parallel b}/\Delta _{Y\parallel a}$~$\geq $~$ \Gamma $.
Therefore, the observed anisotropy of $\Gamma \cong 1.22$ represents
a lower limit for this gap ratio, which is valid for extremely small
ratios $\Delta _{Y}/\Delta _{Nb} \ll 0.1$. For increasing ratio
$\Delta _{Y}/\Delta _{Nb},$ the value $\Gamma ~\cong ~1.22$ requires
a rapid increase of $\Delta _{Y\parallel b}$/$\Delta _{Y\parallel
a}$. In this estimate, $\Gamma $ depends only on the gap ratios and
on the Au/Nb interface transparency.

The anisotropic gap in YBa$_{2}$Cu$_{3}$O$_{7}$ depends on the
in-plane angle $\theta $ ($0$ to $2\pi $), and the angle
$\eta$ (-$\frac{\pi}{2}$ to $\frac{\pi}{2}$) with the
$ab$-plane. Various possible symmetry functions exist to describe this gap.
Here, we consider the following 3D gap
function in YBa$_{2}$Cu$_{3}$O$_{7}$ consisting of a
dominant $d_{x^{2}-y^{2}}$-wave component with an isotropic and an
anisotropic $s$-wave admixture:
\begin{equation}
\Delta _{Y}=\Delta _{Y_{0}}\cos^{2}(\eta )\sum_{i=0}^{2}c_{i}
{\{}\cos^{2}(\theta )-\sin^{2}(\theta ){\}}^{i}
\end{equation}%
with the coefficients $c_{1}>c_{0}, c_{2}$ and $c_{1}+c_{0}+c_{2}=1$.
Here $\Delta _{Y_{0}}$ denotes the magnitude of the YBa$_{2}$%
Cu$_{3}$O$_{7}$ gap at the interface. Consistent with our earlier estimate
$\Delta _{Y\parallel b}/\Delta _{Y\parallel a}>1.22$, the gap-ratio is taken
as $\Delta _{Y\parallel b}/\Delta _{Y\parallel a}\cong 1.5$ in agreement both
with the observed node positions and ARPES.\cite{Aet:2001} With this, the
coefficients are found from the fit to the data $c_{0}=0.15$, $c_{1}=0.83$
and $c_{2}=0.02$. Other choices for the gap symmetry functions lead to
slightly different numbers but will not alter the basic
results of our calculations.

A series of fits is presented in Fig.~\ref{fig3}h: the wider the
tunneling cone (smaller $\kappa$), the smaller is the width of the
oscillations in the $I_{c}R_{n}(\theta ) $-dependence (arrow). The
effective YBa$_{2}$Cu$_{3}$O$_{7}$ gap $\Delta _{Y_{0}}$ and $\gamma
_{B_{Au/Nb}}$ must then become larger. This dependence is presented
in Fig.~\ref{fig3}g. The minimum value $\Delta _{Y_{0}}$ occurs for
normal-incidence tunneling ($\lim \kappa \rightarrow \infty )$, so that $\Delta _{Y_{0}}$~$%
\geq $~6.4~meV. For reasonable $\Delta _{Y_{0}}$ values ($<$~0.5~eV), $%
\gamma _{B_{Au/Nb}}$ varies from 13.6 to about 18. This gives an estimate
for the Au/Nb interface resistance: $R_{B}A\cong 0.36\pm 0.05~n\Omega $~cm$%
^{2}$. In contrast to $\gamma _{B_{Au/Nb}}$, it is not possible
to give an accurate estimate for the YBa$_{2}$Cu$_{3}$O$_{7}$
gap from our data. Therefore, we choose to fit the data with
$\Delta _{Y_{0}}=44$~meV, $\gamma _{B_{Au/Nb}}=16.8$ in Figs.~\ref{fig3}c
and \ref{fig3}f. These are not claimed to be the
correct values; the simulation demonstrates however that large $\Delta
_{Y_{0}}$ may well be consistent with small $I_{c}R_{n}$ values. For the
untwinned case $\kappa=18.4$, corresponding to a tunnel cone with
a full-width-half-maximum (FWHM) of $31.0^{\circ}$ (cosine-term of $D$ not
included). For the twinned case, the $I_{c}R_{n}(\theta )$-dependence is
simulated with the same parameters, except for a slightly smaller cone ($%
\kappa =26.3$, FWHM $=26.0^{\circ}$), and assuming equal presence of both
twin orientations: $\frac{1}{2}[I_{c}R_{n}(\theta )+I_{c}R_{n}(\theta +\frac{\pi}{2})]$. The
smaller tunnel cone for the twinned case is consistent with higher $R_{n}A$%
- and lower $J_{c}$-values. This may result from a slightly thicker tunnel
barrier at the YBa$_{2}$Cu$_{3}$O$_{7}$/Au interface, e.g., due to minor
variations in the Au PLD-conditions, modifying this interface.

The $R_{n}A(\theta )$-dependence is fitted with an ellipsoidal relation of
the conductivity projections along the main crystal directions of the YBa$%
_{2}$Cu$_{3}$O$_{7}$. Written in terms of the $R_{n}A$-values along these
directions, this gives $R_{n}A(\theta )=\sqrt{R_{n}A_{\parallel
b}^{2}\cos^{2}(\theta )+R_{n}A_{\parallel a}^{2}\sin^{2}(\theta )}$. Fig.~\ref{fig3}e shows
the result using $R_{n}A_{\parallel a}=44~n\Omega $~cm$^{2}$ and $R_{n}A_{\parallel
b}=29~n\Omega $~cm$^{2}$. For the twinned case, a geometrical average of the
conductivities is assumed, $R_{n}A(\theta )=2/[R_{n}A^{-1}(\theta
)+R_{n}A^{-1}(\theta +\frac{\pi}{2})]$. The used values in Fig.~\ref{fig3}b read $%
R_{n}A_{\parallel a}=141~n\Omega $~cm$^{2}$, $R_{n}A_{\parallel b}=47~n\Omega $%
~cm$^{2}$. Although these phenomenological fits are indicative,
angle-resolved calculations including aspects of the YBa$_{2}$Cu$_{3}$O$_{7}$
band-structure and band-bending effects are needed for a detailed
understanding. Finally, the $J_{c}(\theta )$ fits are obtained with the ratios of the
$I_{c}R_{n}(\theta )$- and the $R_{n}A(\theta )$-dependencies, the ensemble
of which gives a consistent simulation of the angle-resolved junction properties.

The experimental results support theories based on a 2-band model of the
chains and planes \cite{Mazin:1995,Donovan:1997} modeled with a symmetric,
anti-symmetric and isotropic component. Furthermore, our findings agree with
$c$-axis tunneling from two twinned YBa$_{2}$Cu$_{3}$O$_{7}$ grains to a
Pb counter-electrode that depends on the magnetic field orientation,\cite%
{Kouznetsov:1997} and angle-dependence studies on grain boundary junctions.%
\cite{Lombardi:2002} For all-high-$T_{c}$ junctions and circuits, we mark
the anisotropy as a possible intrinsic source of their limited
reproducibility: both twin orientations may not be uniformly present,
yielding an important variation in $J_{c}$. Control over the crystal
orientation then presents a key to improvement. Another important aspect
concerns the nodes at $5^{\circ}$ from the $\langle 1\!\; 1\!\; 0\rangle $ crystal
direction. The best choice for the electrode-orientation of devices aiming a
$d$-wave induced second harmonic in the current-phase relation, such as
$\frac{\pi}{2}$-SQUIDs based on grain boundary junctions, may therefore deviate
from the $\langle 1\!\; 1\!\; 0\rangle$ crystal direction.

In conclusion, an angle-resolved electron tunneling study using Josephson
junctions with an untwinned YBa$_{2}$Cu$_{3}$O$_{7}$ base-electrode is presented.
Evidence for significant in-plane anisotropy in the electronic properties of
YBa$_{2}$Cu$_{3}$O$_{7}$ is found.

The authors thank M. Yu. Kupriyanov, J. R. Kirtley, C. C. Tsuei, C. W.
Schneider and J. Mannhart for valuable discussions. This work is supported by the Dutch
Foundation for Research on Matter (FOM), and the Netherlands Organization
for Scientific Research (NWO).

\end{document}